\documentclass[11pt]{article}
\usepackage{mathrsfs}
\usepackage[centertags]{amsmath}
\usepackage{amsfonts}
\usepackage{amssymb}
\usepackage{amsthm}
\usepackage{newlfont}
\usepackage[active]{srcltx} % SRC Specials for DVI Searching
\usepackage{eufrak}
\DeclareTextFontCommand{\textfrak}{\frakfamily}
\theoremstyle{definition}
 
 \theoremstyle{remark}
 
 \numberwithin{equation}{section}
\begin{document}

\begin{center}

 \noindent {\LARGE\textbf{New MDS Euclidean and Hermitian self-dual codes over  finite fields }}
\bigskip\bigskip

  \noindent {\Large  Hongxi Tong$^1$\ \ Xiaoqing Wang$^2$}
\medskip

 \noindent{\small  Department of
Mathematics,   Shanghai University, Shanghai 200444.$^{1\ 2}$ \\
(email:  tonghx@shu.edu.cn$^1$\ \ 2625453656@qq.com$^2$)}

\end{center}

 \noindent\textbf{Abstract:} In this paper, we construct MDS Euclidean self-dual codes which are extended cyclic duadic codes.
 And we obtain many new MDS Euclidean self-dual codes. We also  construct MDS Hermitian self-dual codes from generalized Reed-Solomon codes
 and constacyclic codes. And we give some  results on Hermitian self-dual codes, which are the extended cyclic duadic codes.


\section{Introduction}

Let $\mathbb{F}_q$ denote a finite field with $q$ elements. An $[n, k, d]$ linear code $C$ over $\mathbb{F}_q$ is a $k-$dimensional subspace of $\mathbb{F}_q^n$.  These parameters $n$, $k$ and $d$ satisfy $d\leq n-k+1$. If $d=n-k+1$, $C$ is called a maximum distance separable (MDS) code. MDS codes are of practical and theoretical importance. For examples, MDS codes are related to geometric objects called $n-$arcs.

The Euclidean dual code $C^{\bot}$ of $C$ is defined as $$ C^{\bot}:=\left\{ x\in \mathbb{F}_q^n : \sum_{i=1}^n x_i y_i =0,\ \forall y \in C \right\}.$$
If $q=r^2$, the Hermitian dual code $C^{\bot H}$ of $C$ is defined as $$C^{\bot H}:= \left\{ x\in \mathbb{F}_{r^2}^n : \sum_{i=1}^n x_i y_i^r =0,\ \forall y \in C \right\}.$$
If $C$ satisfies $C=C^{\bot}$ or $C=C^{\bot H}$, $C$ is called Euclidean self-dual or Hermitian self-dual, respectively.  There are many papers discussing Euclidean self-dual codes or Hermitian self-dual codes.$^{[2] [12]}$
If $C$ is MDS and  Euclidean self-dual or Hermitian self-dual, $C$ is called an MDS Euclidean self-dual code or
an MDS Hermitian self-dual code, respectively.
In recent years, study of MDS self-dual codes has attracted a lot of attention.$^{[1] [2] [3] [4] [5] [6] [8] [9]}$
 One of these problems in this topic is to determine existence of MDS  self-dual codes.
 When $2|q$, Grassl and Gulliver completely solve the existence of MDS Euclidean  self-dual codes in [4].
 In [5], Guenda obtain some new MDS Euclidean self-dual codes and MDS Hermitian self-dual codes.
 In [8], Jin and Xing obtain some new MDS Euclidean self-dual codes from generalized Reed-Solomon codes.

 In this paper, we obtain some new Euclidean self-dual codes by studying the solution of an equation in $\mathbb{F}_q$.
 And we generalize Jin and Xing's results to MDS Hermitian self-dual codes. We also construct MDS Hermitian self-dual codes from constacyclic codes.
We discuss  MDS Hermitian self-dual codes obtained from extended  cyclic duadic codes.
We give some corrections of the result on  MDS Hermitian self-dual codes obtained from extended  cyclic duadic codes in $[5]$.

\section{MDS Euclidean Self-Dual Codes}

A cyclic code $C$ of length $n$ over $\mathbb{F}_q$ can be considered as an ideal, $<g(x)>$, of the ring $R=\frac{\mathbb{F}_q[x]}{x^n-1}$,
where $g(x)| x^n-1$ and $(n, q)=1$.  The set $T=\{0 \leq i \leq n-1 | g(\alpha^i)=0 \}$ is called the defining set of $C$, where $\mbox{ord}\alpha=n$.

Let $S_1$ and $S_2$ be unions of cyclotomic classes modulo $n$,
such that $S_1\cap S_2= \emptyset$ and  $S_1 \cup S_2=\mathbb{Z}_n\setminus\{0\}$ and $a S_i(\bmod n)=S_{i+1(\bmod 2)}$.
Then  the triple $\mu_a$, $S_1$ and $S_2$  is called a splitting modulo $n$.
Odd-like codes $D_1$ and $D_2$ are  cyclic codes over $\mathbb{F}_q$ with defining sets $S_1$ and $S_2$, respectively.
 $D_1$ and $D_2$ can be denoted by $\mu_a(D_i)=D_{i+1 (\bmod 2)}$.
 Even-like duadic codes $C_1$ and $C_2$ are  cyclic codes over $\mathbb{F}_q$
  with defining sets $\{0\}\cup S_1$ and $\{0\}\cup S_2$, respectively.
  Obviously, $\mu_a(C_i)=C_{i+1(\bmod 2)}$.
  A duadic code of length $n$ over $\mathbb{F}_q$ exists if and only if $q$ is a quadratic residue modulo $n$.$^{[11]}$

Let $n|q-1$ and $n$ be an odd integer.  $D_1$ is a cyclic code with  defining set $T=\left\{1, 2, \cdots ,\frac{n-1}{2}\right\}$.
Then $D_1$ is an $\left[ n, \frac{n+1}{2}, \frac{n+1}{2} \right]$  MDS code.
Its dual $C_1=D_1^{\bot}$ is also cyclic with defining set $T \cup \{0\}$.
There are a  pair of odd-like duadic codes $D_1=C_1^{\bot}$ and $D_2=C_2^{\bot}$ and a pair of even-like duadic codes $C_2=\mu_{-1}(C_1)$.

\textbf{Lemma 1}$^{[5]}$ Let $n|q-1$ and $n$ be an odd integer.
There exists a pair of MDS codes $D_1$ and $D_2$ with parameters $\left[ n, \frac{n+1}{2}, \frac{n+1}{2} \right]$, and $\mu_{-1}(D_i)=D_{i+1 (\bmod 2)}$.

\textbf{Lemma 2}$^{[7]}$  Let $D_1$ and $D_2$ be a pair of odd-like duadic codes of
length $n$ over $\mathbb{F}_q$, $\mu_{-1}(D_i)=D_{i+1 (\bmod 2)}$. Assume that
$$1+ \gamma^2 n=0 \eqno(*)$$
has a solution in  $\mathbb{F}_q$. Let $\widetilde{D}_i=\left\{\widetilde{c}| c\in D_i\right\}$ for $1\leq i\leq 2$ and $\widetilde{c}=(c_0, c_1, \cdots, c_{n-1}, c_{\infty})$ with $c_{\infty}=-\gamma \sum_{i=0}^{n-1}c_i$. Then $\widetilde{D}_1$ and $\widetilde{D}_2$ are Euclidean self-dual codes.

In [7], the solution of $(*)$ is discussed when $n$ is an odd prime. In [5], the solution of $(*)$ is discussed when $n$ is an odd prime power.
 Next, we discuss the solution of $(*)$ for any odd integer $n$ with $n | q-1$.

\textbf{Definition 1 (Legendre Symbol)}$^{[10]}$  Let $p$ be an prime and $a$ be an integer.
$$\left(\frac{a}{p}\right)=\left\{ \begin{array}{cl}
                                     0, & \mbox{if}\ a\equiv 0(\bmod p), \\
                                     1, & \mbox{if}\  a(\neq 0)\  \mbox{is a quadratic residue modulo}\  p, \\
                                     -1, & \mbox{if}\  a\  \mbox{is not a quadratic residue modulo}\  p. \\
                                   \end{array}
                                  \right.$$

\textbf{Proposition 1}$^{[10]}$
$$\left(\frac{a}{p}\right)=\left(\frac{p_1}{p}\right)\cdots \left(\frac{p_s}{p}\right),$$
where $a=p_1 \cdots p_s$.

\textbf{Definition 2 (Jacobi Symbol)}$^{[10]}$  Let $m$ and $n (\neq 0)$ be two integers.
$$\left(\frac{m}{n}\right)=\left(\frac{m}{p_1}\right) \cdots \left(\frac{m}{p_h}\right),$$
where $n=p_1 \cdots p_h$.

We cannot obtain $m (\neq 0)$ is a quadratic residue modulo $n$ from $\left(\frac{m}{n}\right)=1$. But we have the next proposition.

\textbf{Proposition 2 } Let $m(\neq 0)$ and $n$ be two integers and $(m, n)=1$. If $m$ is a quadratic residue modulo $n$, then
$$\left(\frac{m}{n}\right)=1.$$
If $$\left(\frac{m}{n}\right)=-1,$$  then $m$ is not a quadratic residue modulo $n$.

Proof\ \ Obviously.

\textbf{Lemma 3 (Law of Quadratic Reciprocity)}$^{[10]}$   Let $p$ and $r$ be odd primes, $(p, r)=1$.
$$\left( \frac{p}{r} \right) \left(\frac{r}{p}\right)=(-1)^{\frac{r-1}{2}\cdot \frac{p-1}{2}}.$$

\textbf{Corollary 1}  Let $p$ and $r$ be odd primes.

(1) When $p\equiv 1 (\bmod 4)$ or $r \equiv 1 (\bmod 4)$,
$$\left(\frac{p}{r}\right)= \left(\frac{r}{p}\right).$$

(2) When  $p\equiv r\equiv 3 (\bmod 4)$,
$$\left(\frac{p}{r}\right)=- \left(\frac{r}{p}\right).$$

\textbf{Theorem 1}  Let $q=r^t$  and $r$ be an odd prime. Let $n | q-1$ and $n$ be an odd integer.
And $$n=p_1^{e_1} \cdots p_s^{e_s} p_{s+1}^{e_{s+1}} \cdots p_h^{e_h},$$
where
$$p_1 \equiv \cdots \equiv p_s \equiv 3 (\bmod 4), \ \ p_{s+1}\equiv \cdots \equiv p_h \equiv 1(\bmod 4).$$

(1)  When $q\equiv 1 (\bmod 4)$, there is a solution to $(*)$ in $\mathbb{F}_q$.

(2) Let $q\equiv 3 (\bmod 4)$.
If $\sum_{i=1}^s e_i$ is an odd integer, there is a solution to $(*)$ in $\mathbb{F}_q$.

Proof\ \ (1)\  $q \equiv 1(\bmod 4)$.

(1.1) $r \equiv 3 (\bmod 4)$. So we have that $t$ is even. Then every quadratic equation with
coefficients in $\mathbb{F}_r$, such as Eq. $(*)$, has a solution in $\mathbb{F}_{r^2} \subseteq \mathbb{F}_q$.

(1.2) $r \equiv 1 (\bmod 4)$ and $2|t$. The proof is similar as (1.1).

(1.3) $r \equiv 1 (\bmod 4)$ and $2\nmid t$.
$$ 1=\left(\frac{q}{n}\right)=\left(\frac{r}{n}\right)=\left(\frac{r}{p_1}\right)^{e_1}\cdots \left(\frac{r}{p_h}\right)^{e_h}=\left(\frac{p_1}{r}\right)^{e_1}\cdots \left(\frac{p_h}{r}\right)^{e_h}=\left(\frac{n}{r}\right).$$
So $n$ is a quadratic residue  modulo $r$. And $-1$ is a quadratic residue modulo $r$.  So
there is a solution to $(*)$ in $\mathbb{F}_q$.

(2) $q\equiv 3 (\bmod 4)$. Then $r \equiv 3 (\bmod 4)$ and $t$ is odd.
\begin{multline*}
   1=\left(\frac{q}{n}\right)=\left(\frac{r}{n}\right)=\left(\frac{r}{p_1}\right)^{e_1}\cdots \left(\frac{r}{p_s}\right)^{e_s}\left(\frac{r}{p_{s+1}}\right)^{e_{s+1}}\cdots \left(\frac{r}{p_h}\right)^{e_h}\\ =(-1)^{e_1}\left(\frac{p_1}{r}\right)^{e_1}\cdots(-1)^{e_s}
\left(\frac{p_s}{r}\right)^{e_s}\left(\frac{p_{s+1}}{r}\right)^{e_{s+1}}\cdots \left(\frac{p_h}{r}\right)^{e_h}
\\=(-1)^{\sum_{i=1}^s e_i}\left(\frac{p_1}{r}\right)^{e_1}\cdots
\left(\frac{p_s}{r}\right)^{e_s}\left(\frac{p_{s+1}}{r}\right)^{e_{s+1}}\cdots \left(\frac{p_h}{r}\right)^{e_h}
 =(-1)^{\sum_{i=1}^s e_i}\left(\frac{n}{r}\right).
\end{multline*}
If $\sum_{i=1}^s e_i$ is odd, $n$ is not a quadratic residue modulo $r$. And $-1$ is not a quadratic residue modulo $r$. So $-n$ is a quadratic residue modulo $r$. There is a solution to $(*)$ in $\mathbb{F}_q$.

\textbf{Remark} In fact, $n| q-1$, and $n$ is an odd integer and $q\equiv 3 (\bmod 4)$. We can easily prove that there is a solution to $(*)$
 in $\mathbb{F}_q$  if and only if $\sum_{i=1}^s e_i$ is an odd integer.

Let $n|q-1$, $q\equiv 1 (\bmod n).$ $q$ is a quadratic residue modulo $n$. $y^2\equiv q (\bmod n)$.
 Let $q=r^t$ and $q\equiv 3 (\bmod 4)$, where $r$ is a prime.
Then  $r\equiv 3 (\bmod 4)$ and $t$ is odd. Eq. $(*)$ has solutions in $\mathbb{F}_q$  if and only if  Eq. $(*)$ has solutions in $\mathbb{F}_r$. 
And $r$ is a quadratic residue modulo $n$. $(y r^{- \frac{t-1}{2}})^2 \equiv r (\bmod n)$. Let $p$ be an odd prime divisor of $n$.
$r$ is a quadratic residue modulo $p$. Then $\left(\frac{r}{p}\right)=1$. By Law of Quadratic Reciprocity, $p|n$,
$$\left(\frac{p}{r}\right)=\left\{
                             \begin{array}{cc}
                               1,  & p \equiv 1 (\bmod 4) \\
                               -1, & p \equiv 3 (\bmod 4) \\
                             \end{array}
                           \right..$$
The Legendre symbol                           
 \begin{multline*}
    \left(\frac{-n}{r}\right)
=\left(\frac{-1}{r}\right)\left(\frac{p_1}{r}\right)^{e_1} \cdots \left(\frac{p_s}{r}\right)^{e_s}
\left(\frac{p_{s+1}}{r}\right)^{e_{s+1}} \cdots \left(\frac{p_h}{r}\right)^{e_h}  \\
  =(-1)^{1+\sum_{i=1}^s e_i}=\left\{
                             \begin{array}{cc}
                               1,  & \sum_{i=1}^s e_i\ \mbox{is\  odd}  \\
                               -1, & \sum_{i=1}^s e_i\ \mbox{is\  even}  \\
                             \end{array}
                           \right.,
 \end{multline*}
where $n=p_1^{e_1} \cdots p_s^{e_s} p_{s+1}^{e_{s+1}} \cdots p_h^{e_h}$,
  $p_1 \equiv \cdots \equiv p_s \equiv 3 (\bmod 4)$ and $p_{s+1}\equiv \cdots \equiv p_h \equiv 1(\bmod 4)$.
 
\textbf{Theorem 2} Let $q=r^t$ be a prime power,  $n| q-1$ and $n$ be an odd integer. 
Then there exists a pair $D_1$, $D_2$ of MDS odd-like duadic codes of length $n$ and $\mu_{-1}(D_i)=D_{i+1 (\bmod 2)}$, 
where  even-like duadic codes are MDS self-orthogonal,  and $T_1=\left\{1, \cdots, \frac{n-1}{2}\right\}$. Furthermore,

(1) If $q=2^t$, then $\widetilde{D}_i$ are $\left[n+1, \frac{n+1}{2}, \frac{n+3}{2}\right]$ MDS Euclidean self-dual codes.

(2) If $q\equiv 1(\bmod 4)$, then $\widetilde{D}_i$ are $\left[n+1, \frac{n+1}{2}, \frac{n+3}{2}\right]$ MDS Euclidean self-dual codes.

(3) If $q\equiv 3(\bmod 4)$ and $\sum_{i=1}^s e_i$ is an odd integer, then $\widetilde{D}_i$ are 
$\left[n+1, \frac{n+1}{2}, \frac{n+3}{2}\right]$ MDS Euclidean self-dual codes, where $n=p_1^{e_1} \cdots p_s^{e_s} p_{s+1}^{e_{s+1}} \cdots p_t^{e_h}$
and $p_1 \equiv \cdots \equiv p_s \equiv 3 (\bmod 4)$, $ p_{s+1}\equiv \cdots \equiv p_h \equiv 1(\bmod 4)$.

Proof\ \  Obviously, $D_i$ are  $\left[n, \frac{n+1}{2}, \frac{n+1}{2}\right]$ MDS odd-like duadic codes.
If there is a solution to $(*)$, we want to prove $\widetilde{D}_i$ are
 $\left[n+1, \frac{n+1}{2}, \frac{n+3}{2}\right]$ MDS Euclidean self-dual codes, and  we only need to prove  that
$$c \in D_i \ \mbox{and} \ wt(c)=\frac{n+1}{2}, \ \mbox{then}\  wt(\widetilde{c})=\frac{n+1}{2}+1.$$
This is equivalent to prove that $c_{\infty} \neq 0$.  It can be  proved similarly by which proved  in [5].

When $q=2^t$, there is a solution to $(*)$ in $\mathbb{F}_{2^t}$, $\widetilde{D}_i$ are
 $\left[n+1, \frac{n+1}{2}, \frac{n+3}{2}\right]$ MDS Euclidean self-dual codes by Lemma 2.

We can obtain (2) and (3) from Theorem 1 and Lemma 2. Theorem 2 is proved.

We list some new MDS Euclidean self-dual codes in the next table.
\begin{center}
\begin{tabular}{|c|l|}
  \hline
  n & q \\  \hline
  4 & $2^2$, 7 \\
  6 & $2^4$, $3^4$ \\
  8 & $2^3$, $3^6$\\
  10 & $2^6$, $5^6$ \\
  12 &  $3^5$ \\
  14  & $2^{12}$, $3^6$     \\
   16 & 31, $31^2$, $31^3$\\
   18 &  $3^{16}$ \\
   20 &  $5^9$ \\
   22 &   $5^6$ \\
   24 &   $3^{11}$\\
   26 &   $7^4$ \\
   28 &   $7^9$\\
   30 &   59 \\
   156 & $5^4$ \\
  \hline
\end{tabular}
\end{center}

\section{MDS Hermitian Self-Dual Codes}

Let  $n \leq q^2$.  We choose $n$ distinct elements $\{\alpha_1, \cdots, \alpha_n\}$ from $\mathbb{F}_{q^2}$ and $n$ nonzero elements $\{v_1, \cdots, v_n\}$ from $\mathbb{F}_{q^2}$. The generalized Reed-Solomon code
$$GRS_k (\alpha, v):= \left\{(v_1 f(\alpha_1), \cdots, v_n f(\alpha_n)): f(x) \in \mathbb{F}_{q^2}[x], \mbox{deg} f(x)\leq k-1\right\}$$
is a $q^2-$ary $[n, k, n-k+1]$ MDS code, where $\alpha=(\alpha_1, \cdots, \alpha_n)$ and $v=(v_1, \cdots, v_n)$.

\textbf{Theorem 3}  Let $n \leq q$ and $2|n$.  Let  $\{\alpha_1, \cdots, \alpha_n\}$ be $n$ distinct elements from
$\mathbb{F}_q (\subseteq \mathbb{F}_{q^2})$ and $u_i=\prod_{1\leq j\leq n, j\neq i}(\alpha_i-\alpha_j)^{-1}$,   $1\leq i \leq n$.
Then there exist $v_i\in \mathbb{F}_{q^2}$ such that $u_i=v_i^2$, for $i=1, \cdots, n$, and the generalized Reed-Solomon code
$GRS_{\frac{n}{2}} (\alpha, v)$ is an $\left[n, \frac{n}{2}, \frac{n}{2}+1 \right]$
MDS Hermitian self-dual code over $\mathbb{F}_{q^2}$, where $\alpha=(\alpha_1, \cdots, \alpha_n)$ and $v=(v_1, \cdots, v_n)$.

Proof\ \  Obviously,   $u_i (\neq 0) \in  \mathbb{F}_q (\subseteq \mathbb{F}_{q^2})$ for  $1\leq i \leq n$.
So there exist $v_i (\neq 0)\in \mathbb{F}_{q^2}$ such that $u_i=v_i^2$ for  $1\leq i \leq n$.
The generalized Reed-Solomon code $GRS_{\frac{n}{2}} (\alpha, v)$ is an $\left[ n, \frac{n}{2}, \frac{n}{2}+1 \right]$
MDS  code over $\mathbb{F}_{q^2}$. For proving the generalized Reed-Solomon code $GRS_{\frac{n}{2}} (\alpha, v)$ is
Hermitian self-dual over $\mathbb{F}_{q^2}$,  we only prove
$$(v_1 \alpha_1^l, \cdots, v_n \alpha_n^l) \cdot (v_1^q \alpha_1^{k q}, \cdots, v_n^q \alpha_n^{k q})=0, \ \   0 \leq l, k\leq \frac{n}{2}-1.  $$

From the choose of $\alpha_i$,  $v_i$ and [8,  Corollary 2.3],
$$(v_1 \alpha_1^l, \cdots, v_n \alpha_n^l) \cdot (v_1^q \alpha_1^{k q}, \cdots, v_n^q \alpha_n^{k q})=(v_1 \alpha_1^l, \cdots, v_n \alpha_n^l) \cdot (v_1 \alpha_1^{k }, \cdots, v_n \alpha_n^{k })=0, \ \   0 \leq l,  k\leq \frac{n}{2}-1.  $$
So the generalized Reed-Solomon code $GRS_{\frac{n}{2}} (\alpha, v)$ is an $\left[ n, \frac{n}{2}, \frac{n}{2}+1 \right]$ MDS Hermitian self-dual code over $\mathbb{F}_{q^2}$.

%\textbf{Corollary 2} Let $n \leq q$ and $2|n$. There exists an  $[n, \frac{n}{2}, \frac{n}{2}+1]$ MDS Hermitian self-dual code over $\mathbb{F}_{q^2}$.

Next we construct MDS Hermitian self-dual codes from constacyclic  codes.

Let $C$ be an $[n, k]$ $\lambda-$constacyclic code over $\mathbb{F}_{q^2}$  and $(n,  q)=1$. $C$ is considered as an ideal, $<g(x)>$,
of $\frac{F_{q^2}[x]}{x^n-\lambda}$, where $ g(x) | (x^n - \lambda)$. Simply,  $C=<g(x)>$.

\textbf{Lemma 4}$^{[12]}$ Let $\lambda \in \mathbb{F}_{q^2}^{*}$, $r= \mbox{ord}_{q^2} (\lambda)$, and $C$ be a $\lambda-$constacyclic code
 over $\mathbb{F}_{q^2}$. If $C$ is  Hermitian self-dual, then $r | q+1$.

\textbf{Lemma 5}$^{[12]}$   Let $n= 2^a n'$ $(a>0)$ and $r= 2 ^b r'$ be integers such that $2\nmid n'$ and $2\nmid r'$. Let $q$ be an odd prime power such that $(n,q)=1$ and $r|q+1$, and let $\lambda \in \mathbb{F}_{q^2}$ has order $r$. Then Hermitian self-dual $\lambda-$constacyclic codes over $\mathbb{F}_{q^2}$ of length $n$ exist if and only if $b>0$ and $q \not \equiv -1 (\bmod 2^{a+b})$.

Let  $r= \mbox{ord}_{q^2} (\lambda)$ and $r| q+1$.
$$O_{r ,n}= \{1+r j| j=0, 1, \cdots, n-1\}.$$
Then $\alpha^i (i \in O_{r, n})$ are  all solutions of $x^n- \lambda=0$ in some extension field of $\mathbb{F}_{q^2}$, where $ \mbox{ord} \alpha=r n$.
$C$ is called a $\lambda-$constacyclic code with defining set $T \subseteq O_{r, n}$, if $$C=<g(x)>\   \mbox{and}\ \ g(\alpha^i)=0,\ \ \forall i \in T.$$

\textbf{Theorem 4} Let $n=2^a n' (a > 0)$ and $r=2^b r' (b>0)$. $r n | q^2-1$. $\lambda \in \mathbb{F}_{q^2}^*$ with $\mbox{ord} \lambda =r$.
$q \not \equiv -1 (\bmod 2^{a+b})$.
  If $r n|2(q+1)$, there exists an MDS Hermitian self-dual code $C$ over $\mathbb{F}_{q^2}$ with length $n$, $C$ is a $\lambda-$constacyclic code with defining set
$$T=\left\{1+r j | 0\leq j \leq \frac{n}{2}-1\right\}.$$

Proof\ \ If $r n| q^2-1$,  $C_{q^2}(i)=\{i\}$, for $i \in O_{r, n}$, where $C_{q^2}(i)$ denote the $q^2$-cyclotomic coset of $i \bmod r n$.
And $|T|=\frac{n}{2}$, $C$ is an $\left[n, \frac{n}{2}, \frac{n}{2}+1 \right]$  MDS $\lambda-$constacyclic code by the BCH bound of constacyclic code.

When $r n | 2 (q+1)$, $q=\frac{r n l}{2} -1$. Because $q \not \equiv -1 (\bmod 2^{a+b})$, $l$ is odd.
\begin{eqnarray*}
  (-q)(1+r j) &=& -q-q r j \\
    &\equiv & 1-\frac{r n l}{2}+r j \\
    &\equiv& 1+ r(\frac{n}{2}+j)(\bmod r n).
\end{eqnarray*}
So $$(-q)T \cap T= \emptyset.$$
$C$ is MDS Hermitian self-dual by the relationship of roots of a constacyclic code and its Hermitian dual code's roots.

\textbf{Remark}  The MDS Hermitian self-dual constacyclic code obtained from Theorem 4
is different with the MDS Hermitian self-dual constacyclic code in [12], because $(q+1,q-1)=2$ for an odd prime power $q$.

If $r=2$, $C$ is negacyclic. Theorem 4 can be stated as follow.

\textbf{Corollary 2} Let $n=2 ^a n' (a\geq 1) $ and $n'$ is odd. Let
$$q\equiv -1 (\bmod 2^a n^{"})\  \mbox{and} \  q \equiv 2^a -1 (\bmod 2^{a+1}),$$
where $n' | n^{"}$ and $n^{"}$ is odd. Then there exists an MDS Hermitian self-dual  code  $C$ of length $n$  which is negacyclic  with defining set
$$T=\left\{1+2 j| j=0, 1, \cdots, \frac{n}{2}-1 \right\}.$$

Especially, when  $a=1$,  Corollary 2 is similar as [5, Theorem 11].

From Theorem 3 and Theorem 4, we obtain the next theorem.

\textbf{Theorem 5} Let $n \leq q+1$ and $n$ be even. There exists an  MDS Hermitian self-dual  code with length $n$ over $\mathbb{F}_{q^2}$.

\section{MDS Hermitian Self-Dual Codes Obtained from Extended Cyclic Duadic Codes}

Let $D$ be an odd-like duadic code. Let $\gamma \in \mathbb{F}_{q^2}$ be a solution to $$1 + \gamma^{q+1} n= 0.$$
 Obviously, the equation always has a solution in $\mathbb{F}_{q^2}$. Let $c=(c_0, c_1, \cdots, c_{n-1})\in D$.  Define
 $$\widetilde{c}=(c_0, c_1, \cdots, c_{n-1}, c_{\infty}), \ \ \mbox{where}\ \ c_{\infty}=-\gamma \sum_{i=0}^{n-1} c_i.$$
 Let $\widetilde{D}=\{\widetilde{c}| c \in D\}$ be the extended code of $D$.

\textbf{Lemma 6}$^{[2]}$ Let $D_1$ and $D_2$ be a pair of odd-like duadic codes of length $n$ over $\mathbb{F}_{q^2}$. If $\mu_{-q}$ gives
the splitting for $D_1$ and $D_2$, then $\widetilde{D}_1$ and $\widetilde{D}_2$ are Hermitian self-dual.

\textbf{Lemma 7}$^{[2]}$ Let $C$ be a cyclic code over $\mathbb{F}_{q^2}$. The extended code $\widetilde{C}$ is Hermitian self-dual
if and only if $C$ is an odd-like duadic code whose splitting is given by $\mu_{-q}$.

\textbf{Lemma 8}$^{[2]}$  Cyclic codes of length $n$ over $\mathbb{F}_{q^2}$ whose extended code is Hermitian self-dual exist
if and only if for every prime $r$ dividing $n$, either $\mbox{ord}_r(q)$ is odd or $\mbox{ord}_r (q^2)$ is even.

In [5], Guenda  give the next theorem.

\textbf{Theorem 6}$^{[5, \mbox{Theorem\ 8}]}$ Let $q=r^t$ be an odd prime power, and $n=p^m\in \mathbb{F}_r$ a divisor of $q^2+1$, where $p^m\equiv 1 (\bmod 4)$.
Then there exists Hermitian self-dual codes over $\mathbb{F}_{q^2}$ which are MDS and extended duadic codes with
the splitting given by $\mu_{-q}$ and with parameters $\left[n+1, \frac{n+1}{2}, \frac{n+3}{2} \right]$.

In the analysis of Theorem 6 in [5], $D_1$ is an $\left[n, \frac{n+1}{2}, \frac{n+1}{2}\right]$ MDS cyclic code with defining set
$$ T = \left\{\frac{n+3}{4}, \cdots, \frac{3n-3}{4}\right\}.$$
And  $D_1$ is considered as an odd-like duadic code, when $n=p^m(\equiv 1 (\bmod 4)).$
Then the code
$$\widetilde{D}_1 =\left\{\widetilde{c}=(c_0, c_1, \cdots, c_{n-1} c_{\infty})| (c_0, c_1, \cdots, c_{n-1})\in D_1, c_{\infty}=-\gamma \sum_{i=0}^{n-1} c_i\right\},$$
where $\gamma \in \mathbb{F}_{q^2}$ is a root of $1 + \gamma^{q+1} n=0$,  is an MDS Hermitian self-dual codes  by Lemma 6.

Sometimes, $n$ and $q$ satisfy  conditions of Theorem 6,  but $D_1$, with defining set  $T$,  is not an odd-like duadic code.
So it can  be  proved by Lemma 7 that $\widetilde{D}_1$ is not  an (MDS) Hermitian self-dual codes.

\textbf{Example 1} Let  $p=5$, $n=5^2$ and $q=7$,
then $n$ and $q$ satisfy  conditions of Theorem 6. $D_1$ is a $\left[25, 13, 13\right]$   MDS  cyclic code  over $\mathbb{F}_{7^2}$ with defining set
$$T =\{7, 8, \cdots, 17, 18\}.$$  And $12\in (-7)T(\bmod 25) \cap T$.  $D_1$ is not an odd-like duadic code with defining set $T$.
 So  it is  proved by Lemma 7 that $\widetilde{D}_1$ is not  an (MDS) Hermitian self-dual code over $\mathbb{F}_{7^2}$.

 \textbf{Example 2} Let  $p=5$, $n=5^2$ and $q=43$,
then $n$ and $q$ satisfy  conditions of Theorem 6. $D_1$ is a $\left[25, 13, 13\right]$   MDS  cyclic code over $\mathbb{F}_{43^2}$ with defining set
$$T =\{7, 8, \cdots, 17, 18\}.$$  And $13 \in (-43)T (\bmod 25) \cap T$.  $D_1$ is not an odd-like duadic code with defining set $T$.
So  it is  proved by Lemma 7 that $\widetilde{D}_1$ is  not an (MDS) Hermitian self-dual code over $\mathbb{F}_{43^2}$.

Let $n| q^2+1$, $n \equiv 1 (\bmod 4)$ and $q$ be an odd prime power.
 $D_1$ is an $\left[n, \frac{n+1}{2}, \frac{n+1}{2}\right]$ MDS cyclic code over $\mathbb{F}_{q^2}$ with defining set
$$ T = \left\{\frac{n+3}{4}, \cdots, \frac{3n-3}{4}\right\}.$$

We want to prove that $\widetilde{D}_1$ is an MDS  Hermitian self-dual code, so we must prove that $D_1$ is an odd-like duadic code.
It is equivalent to prove that $$(-q) T(\bmod n) \cap T =q T(\bmod n )\cap T = \emptyset.$$
Note that $-T(\bmod n)=T$.

Let $n=4 k+1$, $k\geq 1$.  So $$T=\{ k+1, \cdots, 3 k\}.$$
If $q T(\bmod n )\cap T = \emptyset$ and $q T(\bmod n )\cup T=\{1, 2, \cdots, n-1\}$, then
$$q T= \{1, \cdots, k\}\cup \{3 k+1, \cdots, 4 k\}.$$

There is an $\alpha \in T$ such that $q \alpha \equiv 1 (\bmod n)$. And  $q^2 \equiv -1 (\bmod n)$, so $\alpha \equiv -q (\bmod n)$.

We claim that $$q \equiv \frac{n+3}{4} \ \mbox{or}\  \frac{3 n-3}{4} (\bmod n).$$
If not, then $$\frac{n+3}{4} < q (\bmod n) < \frac{3 n-3}{4}\  \mbox{and}\ \frac{n+3}{4} < -q (\bmod n) < \frac{3 n-3}{4}. $$
Note that $\frac{n+3}{4}\equiv - \frac{3 n-3}{4} (\bmod n)$.
So $$(q+1)(\bmod n)\in T,\ \ (-q+1)(\bmod n)\in T.$$
$$q(-q+1)=-q^2+q \equiv (1+q) (\bmod n) \in q T\cap T.$$
It is a contradiction  to $q T\cap T =\emptyset$.

So $$q \equiv \frac{n+3}{4} \ \mbox{or}\  \frac{3 n-3}{4} (\bmod n).$$
And $n| q^2+1$.
$$q^2+1 \equiv \frac{n^2+6 n+9}{16}+1\  \mbox{or}\ \frac{9 n^2 -18 n+9}{16} +1 \equiv 0 (\bmod n).$$
$$n\equiv 1 (\bmod 4),\ \ (16, n)=1.$$
 So $n|25$. $$n=5\  \mbox{and}\  n=25.$$

When $n=5$, $$q\equiv 2 (\bmod 5) \ \mbox{or} \ q\equiv 3 (\bmod 5).$$

When $n=25$, $$q\equiv 7 (\bmod 25) \ \mbox{or} \ q\equiv 18 (\bmod 25).$$
And $n=25$, $q=7 \equiv 7 (\bmod 25)$ and $ q= 43 \equiv 18 (\bmod 25)$ in Example 1 and Example 2.
So when $n=25$, it is impossible that there is an odd prime power $q$,  with $q\equiv 7 (\bmod 25)$ or $q\equiv 18 (\bmod 25)$,
such that $q T\cap T = \emptyset$, where $T=\{7, 8, \cdots, 18\}.$

When $n=5$ and $n|q^2+1$,  it is easily to prove that Theorem 6 is correct, because  $T=\left\{\frac{n+3}{4},\cdots ,\frac{3n-3}{4}\right\}=\{2, 3\}$.

\textbf{Theorem 7}   Let $q=r^t$ be an odd prime power, and $n=5$ is a divisor of $q^2+1$.
Then there exist Hermitian self-dual codes over $\mathbb{F}_{q^2}$ which are MDS and extended duadic codes with
the splitting given by $\mu_{-q}$ and with parameters $\left[6, 3, 4 \right]$.

If we want to obtain more extended cyclic duadic codes over $\mathbb{F}_{q^2}$, which are Hermitian self-dual,
we shall require that $n|q-1$, $(n, q+1)=1$ and $2 \nmid n$ by the BCH bound of cyclic codes and Lemma 8. So we have the next theorem.

\textbf{Theorem 8}  Let $q=r^t$ be an odd prime power, and $n|q-1, (n, q+1)=1$.
Then there exists Hermitian self-dual codes over $\mathbb{F}_{q^2}$ which are MDS and extended duadic codes with
the splitting given by $\mu_{-q}$ and with parameters $\left[n+1, \frac{n+1}{2}, \frac{n+3}{2} \right]$.

Proof\ \ As $n|q-1$ and  $(n, q+1)=1$,  there is a cyclic  MDS $[n, \frac{n+1}{2}, \frac{n+1}{2}]$ code $D$  over $\mathbb{F}_{q^2}$ with  defining set
$T=\{1, 2, \cdots, \frac{n-1}{2}\}$. And
$$(-q)T\equiv (-1)T \equiv \{n-1, n-2,\cdots,  \frac{n+1}{2} \} (\bmod n), \ \ (-q)T \cap T= \emptyset. $$
So $\widetilde{D}$ is an MDS Hermitian  self-dual code over $\mathbb{F}_{q^2}$ with parameters $\left[n+1, \frac{n+1}{2}, \frac{n+3}{2} \right]$.
Theorem 8 is proved.

\section{Conclusion}

In this paper, we obtain many new MDS Euclidean self-dual codes by solving the equation $(*)$ in $\mathbb{F}_q$.
We generalize the work of [8] to MDS Hermitian self-dual codes, and we construct  new MDS Hermitian self-dual codes from constacyclic codes.
We obtain that there exists an  MDS Hermitian self-dual  code with length $n$ over $\mathbb{F}_{q^2}$, where $n \leq q+1$ and $n$ is even.
And we also discuss these MDS Hermitian self-dual  codes, which are extended cyclic duadic codes.
We give these corrections (Theorem 7 and Theorem 8) of Theorem 6 ($[5, \mbox{Theorem\ 8}]$).

\end{document}